\documentstyle[twoside,12pt]{article}
\input psfig.tex
\def\singlespace {\smallskipamount=3pt plus1pt minus1pt
                  \medskipamount=6pt plus2pt minus2pt
                  \bigskipamount=12pt plus4pt minus4pt
                  \normalbaselineskip=12pt plus0pt minus0pt
                  \normallineskip=1pt
                  \normallineskiplimit=0pt
                  \jot=3pt
                  {\def\smallskip {\vskip\smallskipamount}}
                  {\def\medskip   {\vskip\medskipamount}}
                  {\def\bigskip   {\vskip\bigskipamount}}
                  {\setbox\strutbox=\hbox{\vrule
                    height8.5pt depth3.5pt width 0pt}}
                  \parskip 0pt
                  \normalbaselines}

\def\doublespace {\smallskipamount=6pt plus2pt minus2pt
                  \medskipamount=12pt plus4pt minus4pt
                  \bigskipamount=24pt plus8pt minus8pt
                  \normalbaselineskip=24pt plus0pt minus0pt
                  \normallineskip=2pt
                  \normallineskiplimit=0pt
                  \jot=6pt
                  {\def\smallskip {\vskip\smallskipamount}}
                  {\def\medskip   {\vskip\medskipamount}}
                  {\def\bigskip   {\vskip\bigskipamount}}
                  {\setbox\strutbox=\hbox{\vrule
                    height17.0pt depth7.0pt width 0pt}}
                  \parskip 12.0pt
                  \normalbaselines}

\def\halfspace {\smallskipamount=6pt plus2pt minus2pt
                  \medskipamount=12pt plus4pt minus4pt
                  \bigskipamount=24pt plus8pt minus8pt
                  \normalbaselineskip=16pt plus0pt minus0pt
                  \normallineskip=2pt
                  \normallineskiplimit=0pt
                  \jot=6pt
                  {\def\smallskip {\vskip\smallskipamount}}
                  {\def\medskip   {\vskip\medskipamount}}
                  {\def\bigskip   {\vskip\bigskipamount}}
                  {\setbox\strutbox=\hbox{\vrule
                    height17.0pt depth7.0pt width 0pt}}
                  \parskip 12.0pt
                  \normalbaselines}

\def\pprintspace {\smallskipamount=4pt plus1pt minus1pt
                  \medskipamount=9pt plus2pt minus2pt
                  \bigskipamount=16pt plus4pt minus4pt
                  \normalbaselineskip=14pt plus0pt minus0pt
                  \normallineskip=1pt
                  \normallineskiplimit=0pt
                  \jot=4pt
                  {\def\smallskip {\vskip\smallskipamount}}
                  {\def\medskip   {\vskip\medskipamount}}
                  {\def\bigskip   {\vskip\bigskipamount}}
                  {\setbox\strutbox=\hbox{\vrule
                   height9.5pt depth4.5pt width 0pt}}
                  \parskip 0pt
                  \normalbaselines}

\def\folio{\ifnum\pageno=1\nopagenumbers\else\number\pageno\fi}

\def\refitem{\par\noindent\hangindent 20pt}
\def\wisk#1{\ifmmode{#1}\else{$#1$}\fi}
\def\lt     {\wisk{<}}

\def\le     {\wisk{_<\atop^=}}

\def\gsim   {\wisk{_>\atop^{\sim}}}

\def\muK    {\wisk{{\rm \mu K}}}

\def\deg    {\wisk{^\circ}}
\def\ddeg   {\wisk{{\rlap.}^\circ}}

\begin{document}
\pagestyle{plain}
\pprintspace

\large
\begin{center}
Monte Carlo Simulations of Medium-Scale CMB Anisotropy
\end{center}

\medskip
\normalsize
\pprintspace
\noindent
\begin{center}
A.~Kogut\footnotemark[1]$^{,2}$
and 
G. Hinshaw$^1$
\end{center}
\footnotetext[1]{
~Hughes STX Corporation, Laboratory for Astronomy and Solar Physics, 
Code 685, NASA/GSFC, Greenbelt MD 20771. \newline
\indent~$^2$ E-mail: kogut@stars.gsfc.nasa.gov. \newline
}

\medskip
\normalsize
\pprintspace
\begin{center}
Submitted to {\it The Astrophysical Journal Letters} \\
January 26, 1996\\
\end{center}


\medskip
\begin{center}
\large
ABSTRACT
\end{center}

\normalsize
\noindent
Recent detections of cosmic microwave background (CMB) anisotropy 
at half-degree angular scales
show considerable scatter in the reported amplitude
even at similar angular resolution.
We use Monte Carlo techniques 
to simulate the current set of medium-scale CMB observations, 
including all relevant aspects 
of sky coverage and measurement technique.
The scatter in the reported amplitudes is well within the range
expected for the standard cold dark matter (CDM) cosmological model,
and results primarily from the restricted sky coverage of each experiment.
Within the context of standard CDM 
current observations of CMB anisotropy
support the detection of a ``Doppler peak'' in the CMB power spectrum
consistent with baryon density 
$ 0.01 \lt \Omega_b \lt 0.13$ (95\% confidence)
for Hubble constant $H_0 = 50 ~{\rm km~s}^{-1} ~{\rm Mpc}^{-1}$.
The uncertainties are approximately evenly divided between instrument noise
and cosmic variance arising from the limited sky coverage.

\noindent
{\it Subject headings:} 
cosmic microwave background -- nucleosynthesis

\clearpage
\section{Introduction}
The anisotropy of the cosmic microwave background 
probes the distribution of matter and energy 
prior to the epoch of structure formation.
On large angular scales ($\theta \gsim 2\deg$),
CMB anisotropies reflect perturbations larger than the particle horizon
and thus probe the primordial density distribution.
On angular scales smaller than 2\deg,
causal processes become important
and modify the primordial distribution in model-specific ways.
Observations of the CMB at angular scales near 0\ddeg5 
sample these causal processes 
and offer a powerful test of competing models of structure formation.

The {\it Cosmic Background Explorer} 
provides a clean detection of CMB anisotropy on large angular scales,
measuring the amplitude of the fluctuations 
to a significance of 14 standard deviations (Bennett et al.\ 1996).
On smaller angular scales, the situation is less clear.
A number of groups have reported detections of CMB anisotropy
on angular scales 0\ddeg5 to 1\ddeg5 (Figure 1).
A large scatter is evident, even for experiments at similar angular scales.
Does this scatter reflect 
cosmology (e.g. non-Gaussian features in the distribution of CMB anisotropy), 
instrumentation,
or simply sample variance from the small sky coverage 
(typically less than 0.1\% of the celestial sphere)
of an individual measurement?

Scott, Silk, \& White (1995) approach this problem analytically,
fitting a phenomenological approximation of the CMB power spectrum
to recent detections to obtain best-fit shape parameters.
They conclude that the data do not show excessive scatter
and obtain best-fit parameters consistent with
the existence of a ``Doppler peak'' in the power spectrum
at angular scales near 0\ddeg5.
In this {\it Letter} we use Monte Carlo techniques
to explore further the issue of scatter between different experiments
and the degree to which 
instrument noise, calibration error, and sky coverage
may be expected to affect the results.
For specificity, we work in the context of the 
standard cold dark matter cosmology
(density $\Omega_0=1$, 
cosmological constant $\Lambda=0$, and
Hubble constant $H_0 = h ~{\rm km~s}^{-1} ~{\rm Mpc}^{-1}$ 
with $h=0.5$),
although our results may be extended 
to other choices of model parameters.

\section{Simulations}
We represent the CMB temperature field 
at angular position $(\theta, \phi)$
as a sum over spherical harmonic amplitudes,
$T_{CMB} ~= ~\sum_{\ell m} ~a_{\ell m} Y_{\ell m}(\theta, \phi)$.
The amplitudes $a_{\ell m}$ are specified by the cosmological model:
for a scale-free power spectrum $P(k) \propto k^n$
they are random Gaussian variables
with zero mean and variance
\begin{equation}
\langle a_{\ell m}^2 \rangle ~= C_\ell ~= ~{4 \pi \over 5} ~Q^2
~{\Gamma[l+(n-1)/2] ~\Gamma[(9-n)/2] \over \Gamma[l+(5-n)/2] \Gamma[(3+n)/2] },
\label{alm_eq}
\end{equation}
where $Q$ is the ensemble-averaged quadrupole amplitude
(Bond \& Efstathiou 1987).
The matter content of the universe will modify the coefficients $C_\ell$
to increase power at $\ell \sim 200-1000$ and produce the so-called
``Doppler peaks'' in the power spectrum.
In this {\it Letter} we use CDM power spectra (Stompor 1994)
normalized to the 4-year {\it COBE} value $Q = 18 ~\muK$
(Bennett et al.\ 1996),
evaluated over the range $2 \le \ell \le 500$.
An experiment with Gaussian beam dispersion $\sigma_b$
will observe a smoothed temperature distribution
$T_{\rm obs}(\theta,\phi) = \sum_{\ell m} 
~a_{\ell m} Y_{\ell m}(\theta, \phi) 
~\exp(- \frac{1}{2} \ell(\ell + 1) \sigma_b^2)$.
To reduce instrumental artifacts,
experimenters rapidly sweep the antenna beam 
between two or more locations on the sky,
and combine these to form temperature differences $\Delta T$
whose distribution can then be compared to theory.

A single realization of the CMB sky is specified 
by the set of amplitudes $a_{\ell m}$.
We generate 2000 CMB realizations and use each realization
to simulate the current set of medium-scale CMB experiments.
With a fixed set of amplitudes $a_{\ell m}$
we evaluate $T_{\rm obs}$ 
along the scan patterns and with the beam dispersions 
of the following experiments:
ACME/SP94 (Gundersen et al.\ 1995),
ARGO (De Bernardis et al.\ 1994),
IAB (Piccirillo and Calisse 1993),
MAX4 $\gamma$UMi (Devlin et al.\ 1994),
MAX4 $\sigma$Her and $\iota$Dra (Clapp et al.\ 1994),
MSAM 2-beam and 3-beam (Cheng et al.\ 1994),
Python II (Ruhl et al.\ 1995)
and 
Saskatoon '94 (Netterfield et al.\ 1995).
For each experiment, we combine the smoothed temperatures $T_{\rm obs}$
to form the appropriate set of noiseless temperature differences $\Delta T$
(single difference, double difference, and so forth).
We then add random uncorrelated noise to each set of simulated $\Delta T$
and multiply the noisy result by a factor $(1 + \delta G)$
where $\delta G$ is a Gaussian of mean zero and width determined by
the calibration uncertainty of each experiment
(typically, $|\delta G| \sim 0.1$).
We remove the mean value from each noisy simulated experiment
and further remove a linear drift 
for those groups performing such subtraction.
To speed processing, we neglect possible foreground emission
and correlations from atmospheric or detector noise;
in cases where the experimenters observe at multiple frequencies,
we use a single frequency with noise appropriate to the
final CMB weighted mean used by each group.

From each sky simulation we thus generate 
a set of simulated CMB experiments,
measured at the locations and in a similar manner as the actual experiments.
Cosmic variance, instrument noise, and calibration uncertainties are 
automatically accounted for.
We may then compare the scatter of the simulated experiments in each 
realization to the scatter in the actual data.

We quantify the amplitude of the CMB anisotropy 
for each simulated experiment
using the quadrupole normalization $Q_{\rm flat}$,
the equivalent power for a flat power spectrum 
(Eq. \ref{alm_eq} with $n=1$)
fitted to the simulated temperature differences.
For each simulated experiment, we derive $Q_{\rm flat}$
by maximizing the likelihood
$$
{\cal L}(Q_{\rm flat}^{ij}) ~\propto
~\frac{
\exp(- \frac{1}{2} 
~\sum_{\alpha \beta} 
~\Delta T_\alpha^{ij} ~({\bf M}^{-1})_{\alpha \beta}^{i} ~\Delta T_\beta^{ij} )
 }
{\sqrt{ {\rm det}({\bf M}^i) }},
$$
where ${\bf M}(Q_{\rm flat})^i$ is the flat-spectrum covariance matrix 
between pixels $\alpha$ and $\beta$ computed from the simulations,
the index $i$ tracks the different experiments for each realization,
and $j$ tracks the 2000 CMB realizations.
Note that the subtraction of the mean (and sometimes linear drift)
from the noisy $\Delta T$ correlates the pixels;
we account for this by 
simulating noisy flat-spectrum CMB realizations,
computing the mean covariance matrix at different values of $Q_{\rm flat}$,
and inverting the (noisy) matrix ${\bf M}$ separately 
for each $Q_{\rm flat}$ value.

\section{Results}
For a fixed CDM power spectrum,
we generate a set of simulated values $Q_{\rm flat}^{ij}$
for the $ith$ experiment in the $jth$ realization.
We then define a $\chi^2$ statistic 
to quantify the scatter in each realization,
$$
\chi^2_j ~= ~\sum_i 
~(\frac{(Q_{\rm flat}^{ij} - \langle Q_{\rm flat}^i \rangle)}
{\delta Q^i})^2,
$$
where $\delta Q^i$ is the standard deviation of each simulated $Q_{\rm flat}^i$
evaluated across all 2000 realizations.
The set of 2000 simulated $\chi^2$ values 
can then be compared to the $\chi^2$ of the 
actual data calculated in a similar fashion
(for a recent tabulation see Scott, Silk, \& White 1995).

The actual CMB experiments have $\chi^2 = 10.1$ for 10 degrees of freedom
when compared to a CDM model with baryon density fixed at $\Omega_b = 0.05$.
42\% of the $\Omega_b = 0.05$ simulations had larger $\chi^2$, placing the
current set of CMB experiments near the median of the simulated distribution.
There is no evidence for anomalous scatter in recent detections of CMB 
anisotropy at half-degree angular scales.
We repeat the analysis for models with $0 \le \Omega_b \le 0.2$.
A flat ``no-baryon'' universe 
($\Omega_0=1, ~\Omega_b=0, ~h=0.5$) 
has $\chi^2 = 43.0$
and is rejected at more than 5 standard deviations.
CMB anisotropy data 
clearly support the detection of additional power 
at half-degree angular scales
compared to the {\it COBE} normalization at large angular scales.
In the context of CDM with standard recombination,
we minimize $\chi^2$ to obtain a 95\% confidence interval 
for the baryon density
$ 0.01 \lt \Omega_b \lt 0.13$.
Since we have fixed all other cosmological parameters
(curvature, cosmological constant, Hubble constant, recombination),
this range for $\Omega_b$ can not be regarded as definitive;
nevertheless, it is reassuring to find 
the values derived from CMB anisotropy
to be in agreement with standard Big Bang nucleosynthesis
(Walker et al.\ 1991).

To what extent do instrument noise, calibration uncertainty, and limited sky 
coverage contribute to the uncertainty in recent medium-scale CMB measurements?
We repeated the simulations with the same (noiseless) $a_{\ell m}$ coefficients
and varied the instrument noise and calibration errors.  
For noise levels and sky coverage typical of recent medium-scale experiments,
the combined uncertainties in $Q_{\rm flat}$ are in the range 6--12 \muK,
split almost evenly between cosmic variance and instrument noise;
calibration uncertainties are generally a minor contribution.
Cosmic variance scales as the inverse square root of the sky coverage
(Scott, Srednicki, \& White 1994;
Hinshaw, Bennett, \& Kogut 1994)
and places a lower limit to the accuracy achieved at any noise level.
Determination of the CMB anisotropy amplitude to 10\% accuracy
at 0\ddeg5 angular scales,
let alone multi-parameter fits to the power spectrum,
will require observations with significantly greater sky coverage
than is currently achieved.
Assuming noise levels less than 30 \muK ~per field of view,
a 10\% determination of $Q_{\rm flat}$ requires observing
a $30\deg \times 30\deg$ patch on the sky.

\vspace{18 pt}
\noindent
This work was supported in part by NASA RTOP 399-20-61-01.

\clearpage
\begin{center}
\large
{\bf References}
\end{center}

\normalsize
\pprintspace

\refitem
Bennett, C.L., et al.\ 1996, ApJ Letters, submitted

\refitem
Bond, J.R., \& Efstathiou, G.\ 1987, MNRAS, 226, 655

\refitem
Cheng, E.S., et al.\ 1994, ApJ, 422, L37

\refitem
Clapp, A.C., et al.\ 1994, ApJ, 433, L57

\refitem
De Bernardis, P., et al.\ 1994, ApJ, 422, L33

\refitem
Devlin, M.J., et al.\ 1994, ApJ Letters, 430, L1

\refitem
Gundersen, J.O., et al.\ 1995, ApJ, 443, L57

\refitem
Hinshaw, G., Bennett, C.L., \& Kogut, A. 1994, ApJ, 441, L1

\refitem
Netterfield, C.B., Jarosik, N., Page, L., Wilkinson, D., \& Wollack, E.\ 1995,
ApJ, 445, L69

\refitem
Piccirillo, L., \& Calisse P., 1993, APJ, 411, 529

\refitem
Ruhl, J.E., Dragovan, M., Platt, S.R., Kovac, J., \& Novak, G.\ 1995,
ApJ, 453, L1

\refitem
Scott, D., Srednicki, M., \& White, M.\ 1994, ApJ, 421, L5

\refitem
---, Silk, J., \& White, M.\ 1995, Science, 268, 829

\refitem
Stompor, R.\ 1994, A\&A, 287, 693

\refitem
Walker, T.P., Steigman, G., Schramm, D.N., Olive, K.A., \& Kang, H.-S. 1991,
ApJ, 376, 51

\clearpage
\begin{figure}[t]
\psfig{file=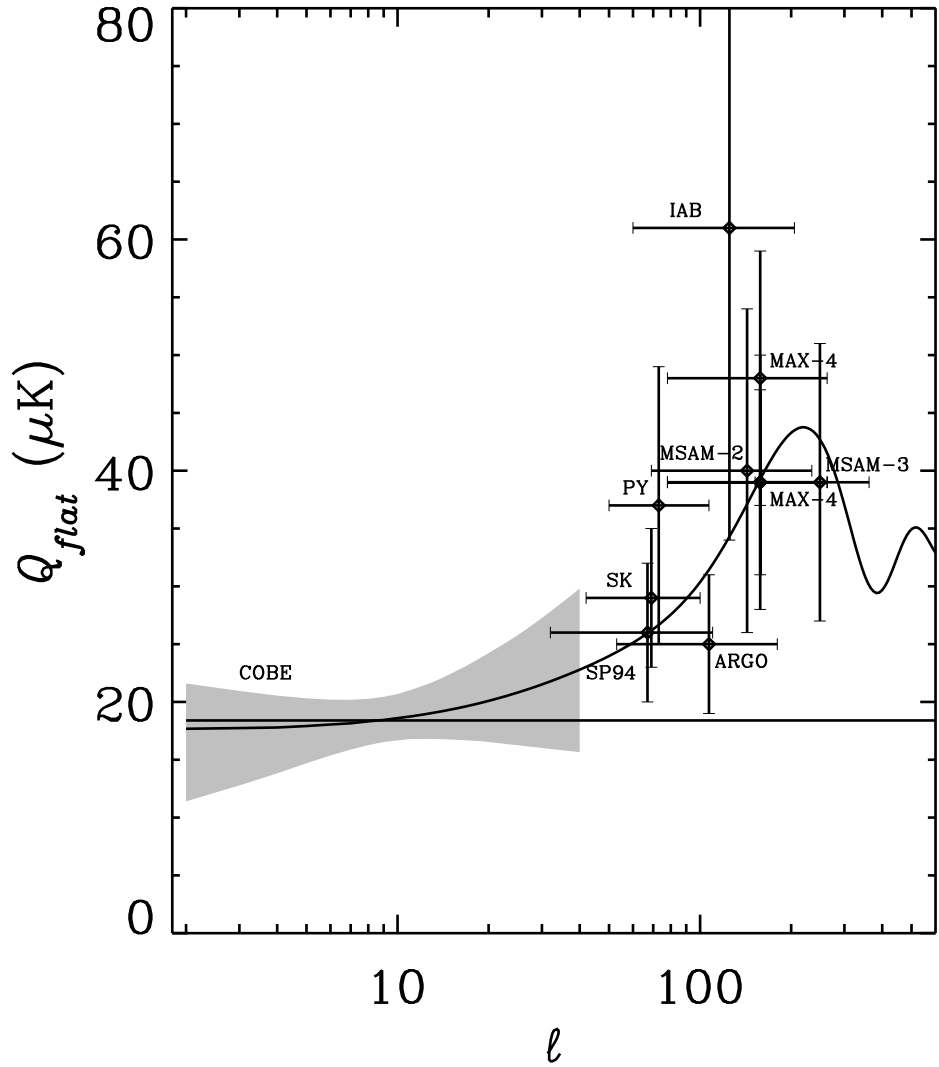,width=6.0in}
\caption{Recent detections of CMB anisotropy,
fitted to the amplitude $Q_{\rm flat}$ of a scale-invariant power spectrum
with power equal to the observed power.
The grey band indicates the 4-year {\it COBE} normalization.
The solid lines show the predicted power spectrum 
for standard CDM (upper curve)
and $\Omega_b = 0$ (lower line).
}
\end{figure}

\end{document}